\documentclass[pra,showpacs]{revtex4}
 \usepackage{amssymb} \usepackage{graphicx} \graphicspath{ {images/} }
  \usepackage{dcolumn} \usepackage{amsmath}
 \usepackage{tikz}
  \usetikzlibrary{shapes,backgrounds}
\usepackage{color}

\begin{document}
\title{Field equations from Killing spinors}
 \author{\"{O}zg\"{u}r A\c{c}{\i}k}
\email{ozacik@science.ankara.edu.tr}
\address{Department of Physics,
Ankara University, Faculty of Sciences, 06100, Tando\u gan-Ankara,
Turkey\\}

\begin{abstract}
From the Killing spinor equation and the equations satisfied by their bilinears we deduce some well known bosonic and fermionic field equations of mathematical physics. Aside from the trivially satisfied Dirac equation, these relativistic wave equations in curved spacetimes respectively are Klein-Gordon, Maxwell, Proca, Duffin-Kemmer-Petiau, K\"{a}hler, twistor and Rarita-Schwinger equations. This result shows that, besides being special kinds of Dirac fermions, Killing fermions can be regarded as physically fundamental. For the Maxwell case the problem of motion is analysed in a reverse manner with respect to the works of Einstein-Groemer-Infeld-Hoffmann and Jean Marie Souriau. In the analysis of the gravitino field a generalised $3-\psi$ rule is found which is termed the vanishing trace constraint.

\end{abstract}
%************************************************************************
\def\nblx#1{\nabla_{X_{#1}}}
\def\sbl#1{(\psi\overline{\psi})_{#1}}

\maketitle
\section{Introduction}
Spinors, with different and more general properties from vectors or tensors, represent fermionic fields or particles on their own \cite{Chevalley, Penrose Rindler I, Penrose Rindler II}. Because one can square them to construct differential forms \cite{Burton,Tucker Rutherford} they also have the power to represent bosonic objects in contrary to tensor fields. Spacetime spinor fields are characterized by the differential equations that they satisfy which are based on the spinor connection that is used for illustrating the dynamics of fermionic fields and the motion of fermionic particles in these fields apriori. Besides many important spacetime spinors such as Dirac, Weyl or Majarona fermions we deduced some of the properties of Killing fermions in our works with physical motivation. Killing fermions are defined as the spinors satisfying the geometric Killing spinor equation and from our point of view are thought as physical more than geometrical. Their geometrical character is basic in theories such as bosonic supergravity with their parametric role for generating supersymmetry transformations and in constructing bosonic supercharges; whereas we focus on their physical properties independent of their partial usage in other physical theories.\\

Other symmetry generating roles of Killing spinors are known where their bilinear covariants were seen to be identical with Killing-Yano forms or closed conformal Killing-Yano forms according to their tensorial degrees and the spinor inner product choosen \cite{Acik Ertem 2015}. Also their original role in the set of specific spinor fields in spacetime was demonstrated in \cite{Acik JMP 2016}. In this work we intend to clarify their physical value by using them to construct many important field equations. After defining the primitive and the principal set of equations that the Killing spinor bilinears satisfy, we will show that these can be used to generate interestingly the Klein-Gordon, Maxwell, Proca, Duffin-Kemmer-Petiau, K\"{a}hler, twistor and Rarita-Schwinger equations in curved spacetimes in a systematic manner.\\

In the analysis of Maxwell equations an important result is uncovered that is reversely parallel to the considerations of Einstein-Groemer-Infeld-Hoffmann and Jean Marie Souriau. Their main motivation was to reach the (passive) equations of motion of test particles from the (active) equations governing the dynamics of the field, whereas our result manifests that the active equations could be deduced from the passive equations at least in the electromagnetic-like interactions. In the case of Rarita-Schwinger field an algebraic constraint appears during the derivation of the resultant field equations and it is observed that this condition is a generalization of the 3-$\psi$ rule, a necessary requirement in supersymmetric theories \cite{Baez and Huerta}. Also, since Rarita-Schwinger and other higher-spin fields carry higher dimensional representations of the Lorentz group or more truly its double cover the spin group, we feel necessary to give the properties of tensor spinors (or spin tensors) preceding the Rarita-Schwinger construction. These representations are irreducible only if some algebraic constraints hold, in our case the constraints will naturally appear and their meanings will be investigated.\\

Killing spinors require curved spacetimes with constant Ricci scalar and their existence endows such spacetimes with a rich physical structure. It is now known that the primitive and principal equation sets belonging to the bilinears of Killing spinors are key for inducing many other bosonic and fermionic fields into spacetime. The understanding of the classical and quantum dynamics of such fields can be contracted to understanding the properties of Killing spinors and their bilinears. Through the point of view of this work, one may see a Killing spinor as a particular kind of Dirac field with many interesting physical properties. So the theory of Killing spinors should be considered as a physical theory. The richness of the theory also provided a route to the arena of higher spin fermionic fields, though we restricted our investigation only to spin-3/2 fields for our current purposes. A more rigorous treatment of higher spin fermionic fields will give a general scheme that will have implications in extended supergravity theories.\\

The Killing fermion sourced electromagnetism in special odd dimensions contains electric, magnetic and composite dyonic monopoles; the motion of charged test objects in these electromagnetic fields are determined by the primitive set of Killing spinor bilinear equations whereas the dynamics of the field is governed by the principal set of equations. Since the latter is a result of the former we concluded that the equation of motions could determine the field dynamics. This result is interesting in itself that more rigorous considerations done in the past follows the reverse path, namely deducing the motions from field equations; the most classical treatments include the names Einstein-Groemer-Infeld-Hoffmann and Jean Marie Souriau. In our analysis of the gravitino field we have reached a condition that has to be satisfied which we called the vanishing trace constraint. This condition contained in particular the $3-\psi$ rule of supersymmetric theories; but since the latter rule is known to be valid only in Minkowski spacetimes with specific dimensions $3,4,6$ and $10$ we concluded that our rule could be checked only in those spacetimes containing at least two Killing spinor fields. Trying to sort out the dimensions of these Lorentzian manifolds will be a problem of its own.\\

The organisation of the paper is as follows. In section II after introducing our notation based on Clifford bundle formulation of physics  and differential geometry \cite{Benn and Tucker,Tucker daktilo}, we give the defining properties of Killing spinors and end up with the introduction of primitive and principal bilinear equations. Section III is devoted to the construction of physical field equations out of the principal equations. In the Maxwell case the generation of the field equations from the equation of motion of test particles is given through the relation between primitive and the principal set. In section IV the Rarita-Schwinger case is analyzed independently because of its different content then the former ones.

\section{Primitive and Principal equations of Killing bilinears}
We assume that the spacetime is a smooth $n$-dimensional real manifold $M$ endowed with two compatible structures: a pseudo-Riemannian metric $g$ with signature $(n_-,n_+)$ and its Levi-Civita connection $\nabla$. Let $C(M,g)$ be the Clifford bundle and $\mathcal{S}(M)$ its spinor bundle considered locally as the minimal left ideal subbundle of the Clifford bundle. If $\psi$ is a spinor and $\overline{\phi}$ a dual spinor defined with respect to a spinor inner product then by Fierz isomorphisms one can write $\psi\otimes\overline{\phi}=\psi \overline{\phi}$. The Clifford product of a spinor $\psi$ and an adjoint spinor $\overline{\phi}$ can be written as an inhomogeneous differential form in terms of projectors $(.)_p$'s onto $p$-form components as follows
\begin{eqnarray}
\psi\bar{\phi}&=&\sum_{p=0}^n(\psi\bar{\phi})_p=\sum_{p=0}^n(\phi,\,e_{I(p)}^{\xi}\psi)e^{I(p)}\\ \nonumber
&=&(\phi,\psi)+(\phi,e_i\psi)e^i+(\phi,e_{ji}\psi)e^{ij}+...+(\phi,e_{i_p...i_2\,i_1}\psi)e^{i_1
i_2...i_p}+...+ (\phi,z^{-1}\psi)z. \nonumber
\end{eqnarray}
 Here $(.,.)$ is the associated inner product on spinors, $e^{i_1i_2...i_p}$ is defined as $e^{i_1}\wedge e^{i_2}\wedge...\wedge e^{i_p}$ which is determined by an arbitrary local co-frame field $\{e^{i}\}$, $\xi$ is the main anti-automorphism which reverses the order of multiplication and $z=*1$ is the volume fixing top-form that is globally defined if $M$ is orientable. Also $1$ is the unit element of the Clifford algebra, $*$ denotes the Hodge map induced by the (smooth) metric tensor field $g$ and multi-index $I(p)$ is a well-ordered $p$-tuple of indices.\\

Killing spinors are defined as the solutions of
\begin{eqnarray}
\nabla_{X}\psi=\lambda \widetilde{X}.\psi
\end{eqnarray}
where $\lambda \in \mathbb{C}$ termed the Killing number which is either real or pure imaginary because of the geometric constraint
\begin{eqnarray}
\mathcal{R}=-4\lambda^{2}n(n-1) \nonumber
\end{eqnarray}
imposed by the existence of such spinor fields in spacetime $(M,g)$ \cite{Kath}. If $\lambda$ is real (imaginary) then the solution is called a real (imaginary) Killing spinor. Here $\widetilde{X}$ is the covector field which is metrically dual to the vector field X and likewise for any multi-vector field $\Omega$, $\widetilde{\Omega}$ will be the corresponding form field.\\

In a recent work \cite{Acik Ertem 2015} we recovered some properties of bilinears generated by twistors and Killing spinors. The Killing spinor case, accompanied by a data set, is more sophisticated and rich. All possible outcomes obtainable from the Killing spinor bilinears is determined by the restrictive reality conditions imposed on them for physical reasons. As a reward we uncovered both kinematical and dynamical equations satisfied by the corresponding generalized Dirac currents of Killing spinors. The \textit{primitive set} of generating equations are seen to be
\begin{equation}
\nabla_{X_a} \sbl p= 2\lambda\, e_a \wedge \sbl {p-1}
\end{equation}
\begin{equation}
\nabla_{X_a} \sbl {p_{*}}= 2\lambda\, i_{X_a} \sbl {p_{*}+1},
\end{equation}
giving rise to the \textit{principal set}
\begin{equation}
d \sbl {p}=0 \qquad,\qquad d^{\dag}\sbl {p}=-2\lambda (n-p+1) \sbl {p-1}\:;
\end{equation}
\begin{equation}
d\sbl {p_{*}}= 2\lambda (p_{*}+1) \sbl {p_{*}+1}\qquad,\qquad d^{\dag}\sbl {p_{*}}=0.
\end{equation}
Here $p_{*}$ means that it has a different parity than $p$, i.e. $p_{*}+p$ is always odd and $0\leq p, p_* \leq n$. Also $\{X_a\}$ is a local frame field satisfying $e^i(X_a)={\delta^i}_a$, $d=e^a\wedge \nabla_{X_a}$ is the exterior derivative and $d^{\dagger}=*^{-1}d*\eta$ is the coderivative where $\eta$ functions as $\eta(\alpha)=(-1)^{p}\alpha$ for a $p$-form $\alpha$.

\section{Generation of physical field equations}

%***********************************************************************

%************************************************************************
\def\nblx#1{\nabla_{X_{#1}}}
\def\sbl#1{(\psi\overline{\psi})_{#1}}

\subsection{Klein-Gordon Field}
Both classical and quantum scalar fields play prominent roles in physics. Scalar fields constitute a first step in flat spacetime field quantization \cite{Ryder}, they do have a well posed Cauchy evolution in globally hyperbolic spacetimes \cite{Wald} and the classical scalar field equation is identical to the equation of quantum motion for a relativistic point particle if written in light-cone coordinates in Minkowski spacetime \cite{Zwiebach}. Though a missing property is that they do not admit any gauge structure.\\

The massive real free scalar field $\varphi$ is taken to satisfy
\begin{eqnarray}
d*d\varphi=m^2 \varphi *1
\end{eqnarray}
where $m$ is the real mass parameter. Our attempt here is to deduce this equation by using spinorial tools. We focus on Killing spinor fields specifically and use their bilinears for this purpose.\\

\textit{\textbf{The construction}}: From our principal equations for degrees $p_*=0$ and $p=1$ and by the abbreviations $S_{\psi}:=(\psi\overline{\psi})_0$ for the scalar part and $\widetilde{V_{\psi}}:=(\psi\overline{\psi})_1$ for the metric dual of the vector part we obtain:
\begin{eqnarray}
  dS_{\psi}&=& 2 \lambda \widetilde{V_{\psi}} \qquad,\qquad d^{\dag}S_{\psi}=0\\
  d\widetilde{V_{\psi}}&=& 0 \qquad,\qquad d^{\dag} \widetilde{V_{\psi}}=-2\lambda n S_{\psi}.
\end{eqnarray}
The second equation in (8) is a trivial one and the second equation in (9) can be rewritten as $d* \widetilde{V_{\psi}}=2\lambda n S_{\psi}*1$. As a last step if we apply the operator $d*$ to the first equation in (8) we get $d*dS_{\psi}=2\lambda d* \widetilde{V_{\psi}}$ which can be reorganised as
\begin{eqnarray}
  d*dS_{\psi}= 4\lambda^2 n S_{\psi}*1
\end{eqnarray}
Note that (10) and (7) become identical if $S_{\psi}:=\varphi$ and $2 |\lambda|\sqrt{n}:=m$. One can easily check the consistency of the reality conditions imposed on $\lambda$, $S_{\psi}$ and $\widetilde{V_{\psi}}$ from pages 9-10 of \cite{Acik Ertem 2015}; the first line of table 3 therein confirms the consistency with reality conditions.
\subsection{Maxwell Field}
Maxwell equations with sources in vacuum or especially in a gravitational field can be written as
\begin{eqnarray}
  d*F&=& j, \\
   dF&=&0
\end{eqnarray}
where $j$ is the current 3-form which has functional dependence on the gravitational field and charged tensor and spinor fields \cite{Benn and Tucker, Baez and Munian}. Although the global topology of the domain where the electromagnetic field is distributed determines whether all closed forms are exact or not, for our local considerations (12) implies that in some neighborhood of every point on $M$ there exists a 1-form $A$ such that $F=dA$. The freedom to choose a potential 1-form $A$ from its equivalence class $[A]_{\sim}:=\{A'\in \Gamma \Lambda^{1}(M)|A'=A+d\lambda; \lambda \in \mathfrak{F}(M)\}$ is known as the local electromagnetic gauge invariance whereas the (Abelian) gauge group is $U(1)$. All electrically charged fields and their $U(1)$ covariant derivatives belong to some representation of $U(1)$; for example the $U(1)$ covariant exterior derivative for a charged scalar field $\phi$ is $$\mathcal{D}\phi=d\phi+iA\phi$$ where $j=Im(\phi*\mathcal{D}\phi^{*})$.\\

\textit{\textbf{The construction}}: Now the first couple of the principal set i.e. equations (5) give rise to
\begin{eqnarray}
d*(\psi\overline{\psi})_{p}=-2 (-1)^{p}\lambda (n-p+1)*(\psi\overline{\psi})_{p-1}\quad,\quad d(\psi\overline{\psi})_{p}=0
\end{eqnarray}
and the second couple of equations given by (6) can be rewritten as
\begin{eqnarray}
d(\psi\overline{\psi})_{p_*}=2\lambda (p_*+1)(\psi\overline{\psi})_{p_*+1}\quad,\quad d*(\psi\overline{\psi})_{p_*}=0.
\end{eqnarray}
If we define $$F_{(p)}:=(\psi\overline{\psi})_{p}\quad,\quad j_{(n-p+1)}:=-2 (-1)^{p}\lambda (n-p+1)*(\psi\overline{\psi})_{p-1}$$ and $$F_{(n-p_*)}:=(\psi\overline{\psi})_{p_*}\quad,\quad j_{(p_*+1)}:=2\lambda (p_*+1)(\psi\overline{\psi})_{p_*+1}$$
then (13) and (14) respectively become
\begin{eqnarray}
d*F_{(p)}=j_{(n-p+1)}\quad,\quad dF_{(p)}=0
\end{eqnarray}
and
\begin{eqnarray}
dF_{(n-p_*)}=j_{(p_*+1)}\quad,\quad d*F_{(n-p_*)}=0.
\end{eqnarray}
(15) and (16) are dual equations in the sense that while the dynamical charge $Q_{(p-2)}:=\int_{\Sigma_{(n-p)}}*F_{(p)}$ is a Noether-Maxwell charge, the dynamical charge $P_{(n-p_*-2)}:=\int_{\Sigma_{(n-p_*)}}F_{(n-p_*)}$ is a topological Faraday charge; here $\Sigma_{(k)}$ is a k-chain in $M$. $Q_{(k)}$ is interpreted as the electric charge and $P_{(k)}$ as the magnetic charge localised on a $(k-2)$-brane. When $p_*=p\mp1$ then $n-p_*=p$ if $n=2p\mp1$; so in a $2p\mp1$ dimensional spacetime, to every electric $(p-2)$-brane there is an associated magnetic $(n-p-1)$ and $(n-p-3)$-brane respectively. For example in 5 dimensions electric-magnetic couple of 1-branes and 0-branes show themselves; so one can make up a composite dyonic 1-brane and 0-brane accordingly \cite{Duff}.\\

\textit{\textbf{Equation of motions generating field equations}}: In our analysis the relation between the primitive set and the principal set of equations for the electromagnetic interactions point out an interesting result. This result is closely related to the problem of motion in an electromagnetic field. Einstein,Groemer,Infeld and Hoffmann worked out the problem of motion in a gravitational field. They had deduced that Einstein's geodesic equation of motion for a test particle now came out of the field equations of general relativity \cite{Wheeler empty}. A similar but mathematically more powerful investigation of the same problem for the case of a spinning charged particle in an electromagnetic field was given by J. M. Souriau \cite{Sternberg ap}. Both investigations were trying to reach the motion of ''charged'' matter starting from the dynamical equations of the fields; but we will here follow the reverse path and will show that the dynamical equations of the electromagnetic field follows from the covariant Lorentz force law.\\

A few steps will give the above result. Both for the primitive and principal equations we take $p=2$ and $p_{*}=1$ then from the second equation of the first set and the first equation of the second set we get $\nabla_{V_{\psi}}\widetilde{V_{\psi}}=2 \lambda i_{V_{\psi}}\widetilde{B_{\psi}}$ and $d \widetilde{B_{\psi}}=0, d^{\dag}\widetilde{B_{\psi}}=-2\lambda (n-1) \widetilde{V_{\psi}}$ respectively. If we make some redefinitions $\widetilde{B_{\psi}}:=F$, $2 \lambda:=q/m$ and $-2\lambda (n-1) *\widetilde{V_{\psi}}:=j_{\psi}$ then the first equation is seen to be the covariant form of the Lorentz force law and the following equations as the Maxwell equations:
\begin{eqnarray}
\nabla_{V_{\psi}}V_{\psi}&=& \frac{q}{m}\widetilde{i_{V_{\psi}}F} \\
d*F=j_{\psi} \quad&,&\quad dF=0.
\end{eqnarray}
Since the principal equations are generated by primitive equations it is clear that in our example the dynamics of electromagnetic fields sourced by Killing fermions can be deduced from the motions of test Killing fermions immersed in these fields.

\subsection{Proca Field}
The Proca equation is a manifestation of passing from the massless electromagnetic field to a field of massive vector bosons \cite{Ryder}. The massive real 1-form field $\mathbf{A}$ satisfies
\begin{eqnarray}
d*d\mathbf{A}=-m^2 *\mathbf{A}
\end{eqnarray}
where $m$ is a real non-zero constant. It is clear that this equation can not be written in a gauge covariant manner, an integrability condition arises by applying $*d$ to the above source-free Proca equation which is $$d^{\dagger}\mathbf{A}=0.$$\\

\textit{\textbf{The construction}}: If we consider the principal equations for $p=2$ and $p_{*}=1$ then with $\widetilde{B_{\psi}}:=(\psi \overline{\psi})_2$
\begin{eqnarray}
d\widetilde{B_{\psi}}&=& 0 \qquad,\qquad d^{\dag} \widetilde{B_{\psi}}=-2\lambda (n-1) \widetilde{V_{\psi}}\\
d\widetilde{V_{\psi}}&=& 4 \lambda \widetilde{B_{\psi}} \qquad,\qquad d^{\dag}\widetilde{V_{\psi}}=0.
\end{eqnarray}
By acting $d*$ to the first equation of (21) we get $d*d\widetilde{V_{\psi}}=4 \lambda d*\widetilde{B_{\psi}}$ which can be reorganised as $d*d\widetilde{V_{\psi}}=4 \lambda *d^{\dag} \widetilde{B_{\psi}}$ since $\widetilde{B_{\psi}}$ is a 2-form. Then from the second equation of (20)
we finally have
\begin{eqnarray}
d*d\widetilde{V_{\psi}}=-8\lambda^2 (n-1)*\widetilde{V_{\psi}}.
\end{eqnarray}
(19) and (22) become identical if we set $8\lambda^2 (n-1)=m^2$ and $\widetilde{V_{\psi}}=\mathbf{A}$. The second equation of (21) is surely the integrability condition. $\lambda\rightarrow 0$ limit is an indication of passing from the massive sector to the massless sector.

\subsection{K\"{a}hler Field}
K\"{a}hler equation is an equation for an inhomogeneous complex differential form $\Phi$ on a pseudo-Riemannian manifold which is designed in 1961 as an alternative for the Dirac's spinor wave equation. There were other attempts in order to understand the properties of electrons with tensor equations by Darwin and independently by Landau and Ivanenko in 1928 \cite{Benn and Tucker}. In the absence of electromagnetic interactions this equation takes the form
\begin{eqnarray}
\displaystyle{\not}d\Phi=\mu \Phi ,
\end{eqnarray}
where $\mu$ is the constant mass parameter. In Minkowski spacetime, K\"{a}hler equation breaks into four algebraically identical copies of equations which in some sense are equivalent to the Dirac equation. This equivalence is possible only if a set of four pairwise-orthogonal primitive idempotents, that project an arbitrary element of the Clifford algebra into minimal left ideals, are globally defined and parallel; so in arbitrary curved spacetimes this is a hard problem generally without any solution. However, a powerful inhomogeneous gravitational field would possibly break the degeneracy carried by the K\"{a}hler electrons, producing four distinct beams in the field as opposed to Dirac electrons; this can give meaning to the extra degrees of freedom.

\textit{\textbf{The construction}}:\\
\textbf{Dimension $3$}:
An inhomogeneous real differential form can be constructed by adding $\widetilde{V_{\psi}}$ and $\widetilde{B_{\psi}}$. Applying Hodge-de Rham operator $\displaystyle{\not}d=d-d^{\dagger}$ to this sum gives $\displaystyle{\not}d(\widetilde{V_{\psi}}+\widetilde{B_{\psi}})=\displaystyle{\not}d\widetilde{V_{\psi}}+\displaystyle{\not}d\widetilde{B_{\psi}}$,
the right hand side of which, because of the first equation of (20) and the second equation of (21), reduces to $d\widetilde{V_{\psi}}+d^{\dagger}\widetilde{B_{\psi}}$. Then using the remaining equations in (20) and (21) we find $$\displaystyle{\not}d (\widetilde{V_{\phi}}+\widetilde{B_{\psi}})=4 \lambda\widetilde{B_{\psi}}+2 \lambda (n-1)\widetilde{V_{\psi}}.$$
If $n=3$ then $\widetilde{V_{\psi}}+\widetilde{B_{\psi}}$ satisfies the K\"{a}hler equation
\begin{eqnarray}
%\nonumber % Remove numbering (before each equation)
 \displaystyle{\not}d (\widetilde{V_{\psi}}+\widetilde{B_{\psi}})=4 \lambda(\widetilde{V_{\psi}}+\widetilde{B_{\psi}}),
\end{eqnarray}
with the identifications $\widetilde{V_{\psi}}+\widetilde{B_{\psi}}=\Phi$ and $4\lambda=\mu$. Here it is important to note that in three dimensions semi-spinor representation of the real Clifford algebra is irreducible which is two dimensional; so the complex semi-spinors have 4 real degrees of freedom. In three dimensions $\widetilde{V_{\psi}}$ has 3 and $\widetilde{B_{\psi}}$ has also 3 real degrees of freedom. This 6 degrees of freedom for $\widetilde{V_{\psi}}+\widetilde{B_{\psi}}=\Phi$ can be reduced to 4 by demanding that $\widetilde{V_{\psi}}$ and $\widetilde{B_{\psi}}$ are timelike forms i.e. $g(V_{\psi},V_{\psi})=-1$ and $g_{2}(\widetilde{B_{\psi}},\widetilde{B_{\psi}})=-1$; so the real K\"{a}hler field will be equivalent to a complex Dirac field in 3 dimensions according to our model.\\

\textbf{Specific odd dimensions}:
If we rewrite the principal set for $p+1$ and $p$ respectively then
\begin{eqnarray}
\nonumber % Remove numbering (before each equation)
  d \sbl {p+1} &=& 0 \qquad,\qquad d^{\dag}\sbl {p+1}=-2\lambda (n-p) \sbl {p}\:; \\ \nonumber
  d\sbl {p} &=& 2\lambda (p+1) \sbl {p+1}\qquad,\qquad d^{\dag}\sbl {p}=0.
\end{eqnarray}
Adding $\sbl {p}$ and $\sbl {p+1}$ and applying the Hodge-de Rham operator to this sum, by also using the above equations we get
$$\displaystyle{\not}d(\sbl {p}+\sbl {p+1})=d\sbl {p}-d^{\dag}\sbl {p+1}=2\lambda (p+1) \sbl {p+1}+2\lambda (n-p) \sbl {p}.$$
By choosing $n=2p+1$ we again reach the massive K\"{a}hler equation (23) with $\Phi=\sbl {p}+\sbl {p+1}$ and $\mu=2\lambda (p+1)$.

\subsection{Duffin-Kemmer-Petiau Field}
Duffin-Kemmer-Petiau equations describe a coupled spin-0 and spin-1 system \cite{DKP}. This first order equation is an equation for an inhomogeneous differential form that also iterates to Laplace-Beltrami equation like the K\"{a}hler equation. $\Phi$ satisfies Duffin-Kemmer-Petiau equation if
\begin{eqnarray}
\nonumber % Remove numbering (before each equation)
  d\Phi_{+} - d^{\dagger}\Phi_{-}=\mu \Phi \nonumber
\end{eqnarray}
where $\Phi_{\pm}:=\frac{1}{2}(1\pm \eta \Phi)$ with $\eta$ the main involutary automorphism of the Clifford algebra and $\mu$ the inertial mass. An equivalent redefinition of this equation can be given by the couple
\begin{eqnarray}
%\nonumber % Remove numbering (before each equation)
  d\Phi_{+} &=& \mu \Phi_{-}\:, \\ \nonumber
  d^{\dagger}\Phi_{-} &=& -\mu \Phi_{+}
\end{eqnarray}
with the obvious integrability conditions $d^{\dagger}\Phi_{+}=0$ and $d\Phi_{-}=0.$\\

\textit{\textbf{The construction}}: If we write the principal equations (5) and (6) for degrees $p$ and $p-1$ and reverse their orders we reach
$$d(\psi\overline{\psi})_{p-1}=2 \lambda p(\psi\overline{\psi})_{p}\quad,\quad\delta(\psi\overline{\psi})_{p-1}=0$$ $$\delta(\psi\overline{\psi})_{p}=-2\lambda (n-p+1)(\psi\overline{\psi})_{p-1}\quad,\quad d(\psi\overline{\psi})_{p}=0,$$
and if we rename the homogeneous bilinears as $\Phi_{\pm}=(\psi\overline{\psi})_{p-1}$,$\Phi_{\mp}=(\psi\overline{\psi})_{p}$ we get
\begin{eqnarray}
%\nonumber % Remove numbering (before each equation)
d\Phi_{\pm}&=&\mu \Phi_{\mp}\quad,\quad\delta(\Phi_{\pm})=0\:, \\ \nonumber
  d^{\dagger}\Phi_{\mp} &=& -\mu\Phi_{\pm}\quad,\quad d\Phi_{\mp}=0
\end{eqnarray}
where $\mu=2|\lambda| p$.

\subsection{Twistors}
\textit{\textbf{The construction}}: If $\psi$ is a Killing spinor
\begin{equation}
\nabla_{X}\psi=\lambda \widetilde{X}.\psi
 \end{equation}
then we define the Killing reversal $\psi^{\varsigma}$ of $\psi$ by
\begin{equation}
\nabla_{X}\psi^{\varsigma}=-\lambda \widetilde{X}.\psi^{\varsigma}\;,
\end{equation}
\cite{Acik JMP 2016, Acik CQG}. To every Killing spinor pair $\psi,\psi^{\varsigma}$ there corresponds a twistor pair $\Psi^{+},\Psi^{-}$ such that $$\Psi^{\pm}=\psi\pm\psi^{\varsigma},$$
trivially the Killing reversals of the induced twistors are ${\Psi^{\pm}}^{\varsigma}=\pm\Psi^{\pm}$. This result follows easily as:
$$\nabla_{X_a}\Psi^{\pm}=\nabla_{X_a}(\psi\pm\psi^{\varsigma})=\nabla_{X_a}\psi\pm\nabla_{X_a}\psi^{\varsigma}=\lambda e_{a}.\psi\mp\lambda e_{a}.\psi^{\varsigma}=\lambda e_{a}.\Psi^{\mp},$$ and if we Clifford contract both hand sides from left with $e^{a}$ and use the identity $e^a e_a=n$ then we obtain $$\displaystyle{\not}D\Psi^{\pm}=n\lambda \Psi^{\mp}$$ and if we put the last identity into the last equality of the first relation we reach the desired result
\begin{equation}
\nabla_{X_a}\Psi^{\pm}=\frac{1}{n}e_{a}.\displaystyle{\not}D\Psi^{\pm}\;,
\end{equation}
namely the twistor equation.\\

\section{Rarita-Schwinger Field}
The gauged supersymmetric extension of general relativity is supergravity and the graviton acquires a fermionic partner called the gravitino where the gauge principle is local supersymmetry. Gauge fields of the theory are the moving coframe field $e^a$ that describes gravity coupled to fermions and the spinor-valued 1-form field $\Psi$. The graviton and the gravitino belong to the (2,3/2) representation of the superalgebra including one supercharge. A spin-3/2 particle is the key prediction of supergravity \cite{Freedman Van Proeyen}.\\

\textit{\textbf{Tensor Spinors}}: Independent of the dimension and signature, the higher dimensional half-integer representations of the spin group can be constructed by tensoring the tensors over an inner product space with the spinors of the associated Clifford algebra. These \textit{tensor spinors} could also be thought of as \textit{spinor-valued tensors} \cite{Benn and Tucker}. A first example for tensor spinors can be given by \textit{spinor-valued $1$-forms} over a pseudo-Riemannian spacetime assumed as a spin manifold $(M,g)$. If ${e^a}$ is any co-frame and ${\mathbf{b}_i}$ is any standard spinor frame then one can write $$\Psi=\psi_a \otimes e^a$$ or equivalently $$\Psi=\mathbf{b}_i \otimes \psi^i$$ where the spinors $\psi_a$'s and $1$-forms $\psi^i$'s are defined respectively as $\psi_a=\mathbf{b}_i \psi_a^i$ and $\psi^i=\psi_a^i e^a$.\\

 In four spacetime dimensions the complex Clifford algebra is isomorphic to the algebra of $4\times4$ complex matrices $\mathbb{C}(4)$ and the even algebra is isomorphic to $\mathbb{C}(2)\oplus \mathbb{C}(2)$ whereas the real subalgebra is isomorphic to $\mathbb{R}(4)$. The irreducible representations of these algebras are carried respectively by complex (Dirac) spinors, complex semi-spinors (Weyl/chiral spinors) and real (Majorana) spinors. Since $z^2=-1$ for Lorentzian signature in four dimensions, we can choose $\psi_a$'s as chiral spinors satisfying $iz\psi_a=\psi_a$, hence they carry irreducible representations of the spin group $SL(2,\mathbb{C})$. $1$-forms Clifford anti-commute with the volume form $z$, then they can be thought of as $(1,1)$ tensors on spinors changing the semi-spinor components under their left Clifford action. Consequently we may regard spinor-valued $1$-forms as bidegree (1,2) tensors on spinor fields. If $u$ and $v$ are two chiral spinors lying in the same component with $\psi_a$, then one can define $$\Psi(u,v):=(u,\psi_a)e^a.v .$$ It turns out that irreducible $SL(2,\mathbb{C})$ representations are carried by spinor-valued $1$-forms that are totally symmetric in their covariant and contravariant arguments separately \cite{Penrose Rindler II}. This is equivalent to saying that the spinor-valued $1$-form is an irreducible spin tensor if it is ''traceless'': $$e^a.\psi_a=0.$$ For example if we assume that $\psi$ satisfies the Weyl equation $\displaystyle{\not}D\psi=0$ then $\Psi$ is traceless if $\psi_a:=\nabla_{X_a}\psi.$\\

 The covariant derivative $\nabla_X$ on spinors and tensors can be extended by the Leibniz rule to a covariant derivative on spinor-valued $1$-forms, also denoted by $\nabla_X$, as $$\nabla_X\Psi=\nabla_X \psi_a \otimes e^a+\psi_a \otimes \nabla_Xe^a.$$ A higher half-integral spin representation of Clifford algebra on spinor-valued $1$-forms can be defined by $$a.\Psi:=(a.\psi_b)\otimes e^b.$$ Then with the tracelessness condition the Dirac-like equation $$\displaystyle{\not}D\Psi=m \Psi$$ constitute the spin $3/2$ Rarita-Schwinger equation. A spinor-valued $p$-form can likewise be defined as $$\Psi:=\psi_{I(p)}\otimes e^{I(p)}$$ where $I(p)$ is an ordered multi-index for a $p$-form basis. If $a$ is any $q$-form and $\Psi$ a spinor-valued $p$-form then a spinor-valued $p+q$-form can be build up from them as $$a\wedge \Psi=\psi_{I(p)}\otimes a\wedge e^{I(p)}.$$ The covariant derivative is again extended to this sector by the Leibniz rule and the spinor covariant exterior derivative D maps spinor-valued $p$-forms to spinor-valued $p+1$-forms: $$D\Psi=e^a \wedge \nabla_{X_a} \Psi.$$ If $\{\mathbf{b}_i\}$ is a standard spinor frame and $\{\psi^i=\psi^i_{I(p)} e^{I(p)}\}$ a set of $p$-forms then $\Psi$ may be expanded as $$\Psi=\mathbf{b}_i \otimes \psi^i$$ and equivalently the spinor exterior covariant derivative can be written as $$D\Psi=d\Psi+\omega.\Psi$$ where $d\Psi:=\mathbf{b}_i \otimes d\psi^i$ and $\omega$ is the Clifford-valued connection $1$-form $\omega:=\frac{1}{4}e^{pq} \otimes \omega_{pq}$. If $n^{A(q)}=\sum_{r}(n^{A(q)})_{J(r)} e^{J(r)}$'s are arbitrary Clifford forms, a Clifford-valued $q$-form $N=n^{A(q)} \otimes e_{A(q)}$ left acts on $\Psi$ as $$N.\Psi=n^{A(q)}.\psi_{I(p)} \otimes e_{A(q)}\wedge e^{I(p)}.$$ Another possible description for the Rarita-Schwinger equation can be given as $$e*D\Psi=0$$ where $e:=e^a \otimes e_a$.\\

\textit{\textbf{The construction}}: Now we want to search for a tensor spinor $\Psi$ with two characteristic components $\Psi_{\mathrm{I}}$ and $\Psi_{\mathrm{II}}$ built up from the external tensor product of a set of Killing spinors $\{\psi_a\}$ with the corresponding set of degree 1 generalised Dirac currents $\{\widetilde{V_a}:=(\psi_a\overline{\psi_a})_1\}$ and the set of degree 2 Dirac currents $\{\widetilde{B_a}:=(\psi_a\overline{\psi_a})_2\}$ respectively; here $V_a$ is the vector current and $B_a$ is the bivector current of the Killing spinor $\psi_a$. The construction of the component tensor spinors are
$$\Psi_{\mathrm{I}}:=\sum_a \psi_a \underset{ext}{\otimes} \widetilde{V_a}$$
and
$$\Psi_{\mathrm{II}}:=\sum_a e^b.\psi_a \underset{ext}{\otimes} i_{X_b}\widetilde{B_a}.$$
For convenience we use the same symbol $\nabla$ for all connections including the connection on spinor-valued 1-forms $\nabla^{3/2}$ and the spinor connection $\nabla^{1/2}$ and of course the ordinary Riemannian connection $\nabla$, and we also use $\otimes$ instead of $\underset{ext}{\otimes}$.\\

We will show that the spin-3/2 tensor $\Psi:=\Psi_{\mathrm{I}}-\Psi_{\mathrm{II}}$ is a Rarita-Schwinger field with mass $-\lambda n$ under the assumption of the tracelessness condition or the vanishing trace constraint $$\sum_a \widetilde{V_a}.\psi_a=0$$.\\

First we start by computing
\begin{eqnarray}
 \nonumber % Remove numbering (before each equation)
 \nabla_{X_c}\Psi_{\mathrm{I}} &=& \sum_a \big(\nabla_{X_c}\psi_a \otimes \widetilde{V_a}+\psi_a \otimes \nabla_{X_c}\widetilde{V_a}\big) \\ \nonumber
   &=& \lambda \sum_a \big({e_c}.\psi_a \otimes \widetilde{V_a}+\psi_a \otimes \nabla_{X_c}\widetilde{V_a}\big),  \nonumber
\end{eqnarray}
using (4) in the last term of the right hand side we obtain
\begin{eqnarray}
 %\nonumber % Remove numbering (before each equation)
 \nabla_{X_c}\Psi_{\mathrm{I}} = \lambda \sum_a \big({e_c}.\psi_a \otimes \widetilde{V_a}+2\psi_a \otimes i_{X_c}\widetilde{B_a}\big).
\end{eqnarray}
The second step of the computation is given by
\begin{eqnarray}
 \nonumber % Remove numbering (before each equation)
 \nabla_{X_c}\Psi_{\mathrm{II}} &=& \sum_a \big(\nabla_{X_c}(e^b.\psi_a) \otimes i_{X_b}\widetilde{B_a}+e^b.\psi_a \otimes \nabla_{X_c}i_{X_b}\widetilde{B_a}\big) \\ \nonumber
   &=&\sum_a \big(-{\omega^b}_k(X_c) e^k.\psi_a \otimes i_{X_b}\widetilde{B_a}+\lambda e^b e_c.\psi_a \otimes i_{X_b}\widetilde{B_a}+e^b.\psi_a \otimes i_{X_b}\nabla_{X_c}\widetilde{B_a}+e^b.\psi_a \otimes i_{\nabla_{X_c}X_b}\widetilde{B_a}\big), \nonumber
\end{eqnarray}
and since the first and last terms at the right hand side sum up to zero we get
\begin{eqnarray}
 %\nonumber % Remove numbering (before each equation)
 \nabla_{X_c}\Psi_{\mathrm{II}}= \sum_a \big(\lambda e^b e_c.\psi_a \otimes i_{X_b}\widetilde{B_a}+e^b.\psi_a \otimes i_{X_b}\nabla_{X_c}\widetilde{B_a}\big). \nonumber
\end{eqnarray}
From (3) $\nabla_{X_c}\widetilde{B_a}=2\lambda e_c \wedge \widetilde{V_a}$, so
\begin{eqnarray}
 %\nonumber % Remove numbering (before each equation)
 \nabla_{X_c}\Psi_{\mathrm{II}}= \lambda \sum_a \big( e^b e_c.\psi_a \otimes i_{X_b}\widetilde{B_a}+2e_c.\psi_a \otimes \widetilde{V_a}- 2e^b.\psi_a \otimes e_c \wedge i_{X_b}\widetilde{V_a}\big). \nonumber
\end{eqnarray}
Finally we obtain
\begin{eqnarray}
 %\nonumber % Remove numbering (before each equation)
 \nabla_{X_c}\Psi_{\mathrm{II}}= \lambda \sum_a \big( e^b e_c.\psi_a \otimes i_{X_b}\widetilde{B_a}+2e_c.\psi_a \otimes \widetilde{V_a}- 2\widetilde{V_a}.\psi_a \otimes e_c\big).
\end{eqnarray}
Clifford multiplying (30) and (31) by $e^c$ and using the identities $e^c e_c=n$ and $e^c e^b e_c=(2-n)e^b$ we get
\begin{eqnarray}
 %\nonumber % Remove numbering (before each equation)
\displaystyle{\not}D\Psi_{\mathrm{I}} = \lambda n\Psi_{\mathrm{I}}+2 \lambda \Psi_{\mathrm{II}},
\end{eqnarray}
\begin{eqnarray}
 %\nonumber % Remove numbering (before each equation)
 \displaystyle{\not}D\Psi_{\mathrm{II}}= \lambda (2-n) \Psi_{\mathrm{II}}+2 \lambda n \Psi_{\mathrm{I}}-2 \lambda e^c (\sum_a\widetilde{V_a}.\psi_a)\otimes e_c.
\end{eqnarray}
Subtracting (33) from (32) and using the vanishing trace constraint the result becomes
\begin{eqnarray}
 %\nonumber % Remove numbering (before each equation)
\displaystyle{\not}D\Psi = -\lambda n\Psi.
\end{eqnarray}
$\lambda$ can easily be chosen as a negative real number so that the mass becomes positive. The vanishing trace constraint is a condition which generalizes the $3-\psi$ rule in supersymmetric theories. $3-\psi$ rule requires $\widetilde{V_a}.\psi_a$ for all $a$ and is valid only in $3,4,6$ and $10$ dimensional Minkowski spaces \cite{Baez and Huerta}. So our vanishing trace condition is more general and should be checked for specific spacetimes containing at least two Killing spinors.

 \end{document}